\documentclass[letterpaper,twocolumn,10pt]{article}
\usepackage[T1,hyphens]{url}
\usepackage{./usenix,epsfig,endnotes}
\usepackage{amsmath}
\usepackage{listings}
\usepackage[utf8]{inputenc}
\usepackage[T1]{fontenc}
\usepackage{url}
\usepackage{verbatim}
\usepackage{amssymb}
\usepackage[export]{adjustbox}
\newcommand{\eat}[1]{}
\newcommand{\msg}[1]{{\small \textsf{#1}}} 

\lstdefinestyle{myhtmlstyle}{
basicstyle=\ttfamily\tiny,
breaklines=true
}
\begin{document}

\date{}

\title{\Large \bf Web Infrastructure to Support e-Journal Preservation (and More)}

\author{
{Herbert Van de Sompel}\\
Los Alamos National Laboratory\\
herbertv@lanl.gov \\
http://orcid.org/0000-0002-0715-6126
\and
{David S. H. Rosenthal}\\
LOCKSS Program, Stanford University Libraries\\
dshr@stanford.edu\\
http://orcid.org/0000-0002-3094-1376
\and
{Michael L. Nelson}\\
Old Dominion University\\
mln@cs.odu.edu\\
http://orcid.org/0000-0003-3749-8116
} 

\maketitle
\thispagestyle{empty}

\subsection*{Abstract}

E-journal preservation systems have to ingest millions of articles each
year. Ingest, especially of the ``long tail'' of journals from small publishers,
is the largest element of their cost. Cost is the major reason that archives contain less than half the content
they should. Automation is essential to minimize these costs.
This paper examines the potential for automation beyond the status quo based on
the API provided by CrossRef, ANSI/NISO Z39.99 ResourceSync, 
and the provision of typed links in publishers' HTTP response headers.
These changes would not merely assist e-journal preservation and other
cross-venue scholarly applications, but would help remedy the gap that 
research has revealed between DOIs' potential and actual benefits.

{\tiny Copyright \copyright 2016 Herbert Van de Sompel \& David S. H. Rosenthal \& Michael L. Nelson}

\section{Introduction}
\label{sec:Intro}

A study~\cite{StmReport2015}
estimated that at least 1.8M articles were published in academic journals
in 2012.
Systems that preserve these journals in electronic form must ingest
these articles in real time, so their ingest processes must be highly
automated.
Current systems have implemented two different techniques for doing
so.
\emph{Harvest} uses web archiving technology to collect newly-published
content from the journal publisher's web site,
whereas \emph{File Transfer} relies on the publisher to push such
content to the archive via FTP or rsync.
Portico~\cite{PorticoWebsite} relies on file transfer,
the Global LOCKSS Network~\cite{LOCKSS} uses harvest,
and the CLOCKSS Archive~\cite{CLOCKSS} uses both,

In each case, the automated ingest process has to answer three questions:
\begin{enumerate}
\item \emph{Manifest}: what should be ingested?
\item \emph{Completeness}: was everything that should have been
ingested actually successfully ingested?
\item \emph{Bibliography}: what bibliographic metadata describes the
ingested content? This includes determining the Identity of the object.
\end{enumerate}

Via an organization named CrossRef~\cite{CrossRef},
articles in most journals are assigned a Digital Object Identifier (DOI),
a location-independent identifier that can be,
but alas often is not~\cite{EvanescentWeb},
used to cite the article.
CrossRef has developed an API for querying information about
DOIs~\cite{CrossRefApi}.
This paper examines various possible ways this API,
ANSI/NISO Z39.99 ResourceSync, and the recent ``Signposting'' proposal from Van de Sompel and
Nelson~\cite{vdSompel2015},
could together be used to improve the ingest processes of e-journal
preservation systems.
It concludes with a set of recommendations to CrossRef and publishers,
aimed not merely at assisting e-journal preservation,
and other applications such as citation management,
but also at helping remedy the gap research has revealed between DOIs'
potential and actual benefits~\cite{vdSompel-WWW2016}. As such, this paper calls 
upon CrossRef, academic publishers, and, more broadly, all operators of portals involved in scholarly communication 
to start a discussion about the feasibility of the recommendations, 
and to eventually implement (a version of) them, not just as a means to support archival use cases, but to 
increase overall interoperability for web-based scholarly communication. 
 
\section{Current Harvest Ingest Processes}
\label{Sec:Harvest}

The current harvest content ingest pipeline for the CLOCKSS Archive is
documented in detail in the Archive's TRAC audit materials~\cite{CLOCKSS-Audit}.
Briefly, the process of ingesting journal content involves:
\begin{itemize}
\item At regular intervals a web crawler visits a known ``start page''
for the journal. This is typically the journal's home page or the current volume's
table of contents (ToC) page. It collects the current state of the start page.
\item The crawler identifies as yet unvisited links on the start page
and follows them to collect the linked-to content.
These pages frequently link to many things other than journal content,
so the crawler has heuristics, called crawl rules,
intended to eliminate those links unlikely to point to journal content.
\item The process repeats to find articles.
Typical journals have a structure in which a home page links to volume
ToC pages, which link to issue ToC pages, which link to article ``landing pages'' 
(See Figures~\ref{fig:Patterns1}, and~\ref{fig:Patterns2}).
\item Landing pages typically contain an abstract and metadata for
the article. The crawler collects the landing page and analyzes its links.
Among them are typically full-text HTML and PDF versions,
and any supplemental materials. Landing pages are not universal,
in some cases (e.g., ~\cite{vdSompel2015}) the ToC page links directly to the full-text HTML.
\item The crawler again identifies these links using heuristics,
and follows them to collect the HTML, the PDF and the supplemental materials.
\item The crawler analyzes the links in the HTML.
Again, heuristics are used to identify the link targets that form part of the
article and must thus be saved.
\begin{itemize}
\item Some will not be part of this or any of this journal's other
articles, such as links to articles cited from other journals.
These links should not be followed.
\item Some of their targets will be part of the article,
and not part of any other article, such as the images of figures,
and the spreadsheets that generated graphs. These links should be followed.
\item Some will be to targets shared by many articles in the journal,
an obvious example being the link back to the ToC page.
These links should only be followed if they have not previously been.
\end{itemize}
\item The metadata extractor~\cite{LOCKSS-MetadataExtractor}
analyzes the landing page, the HTML or the PDF,
depending upon where suitable metadata may be found,
and creates an entry containing the metadata in a metadata
database~\cite{LOCKSS-MetadataDatabase}.
This again uses heuristics to locate the metadata in one or more of the
article's files, and to normalize it into a canonical form~\cite{RosenthalMetadataIIPC2013}.
\end{itemize}

In an ideal world, the heuristics work well and this process collects
everything the publisher ``objectively'' published,
in the sense of making it available to readers,
shortly after they did so and obtains useful metadata.
This process works well, especially for journals published on one of the major publishing
platforms, such as HighWire, Atypon, PLOS or Open Journal System.
But in the real world, the heuristics are less than perfect.

Publishing platforms and publishers are continually
``optimizing their user's experience'' by making changes to their
web sites look-and-feel. Some of these changes can render the heuristics ineffective.
The LOCKSS system implements automated ``substance'' monitoring.
The system is configured with per-publisher information describing
the minimum number and size of PDF and HTML files expected from
each journal in a publishing cycle; it alerts if these limits are violated.

Further, these heuristics operate on a single source of data,
the publisher's web site. All large-scale data feeds contain noise,
and this data is no exception.
Examples include aberrant spellings of publisher and journal names,
and malformed, unregistered or incorrect DOIs.

The ingest process thus needs continual monitoring and adjustment
to ensure that the three questions are answered satisfactorily,
and inconsistencies addressed.
This is a significant part of the total cost of ingest,
which is the largest part of the total cost of preservation.
Automation of this monitoring would be greatly aided if there were
a second, independent source of data against which the data extracted from
the publisher's web site could be compared.

To what extent can data obtained from CrossRef play this role?
It is not completely independent, as it also comes from the publisher.
But CrossRef applies some validation techniques to the data it receives,
and the channel by which it is received is quite different,
so it may play a significant role.

\section{Current File Transfer Ingest Processes}
\label{Sec:FileTransfer}

One might expect that the process of ingesting content via file
transfer would be reliable, and would not need heuristics, but this is not the case.
A 2014 study by University of California librarians~\cite{UCPortico}
assessed the reliability of the holdings Portico obtains via file transfer
by examining ``a random sample of 104 titles'' from ``a list of 10,460 CDL
licensed titles from the Portico e-Journal Publishers list'':
\begin{quotation}
The holdings for each title were determined by looking up each title on
the publisher’s website,
recording the volumes,
issues/volume,
year,
month and supplements (if any),
and then totaling the number of issues for each title.
The holdings reported by the publisher were compared to the holdings
currently accessible in the Portico Archive and the percentage of
“preserved” issues/title was calculated.
No attempt was made in this study to ascertain if each issue is complete.
\end{quotation}
The study concluded that:
\begin{quotation}
The major finding is that only 50\% of the issues to which UC currently
has access via the publishers’ sites are accessible now in the
Portico Archive.
\end{quotation}
Portico's response was ``that some lacunae were caused by issues lacking
data needed to properly ingest them;
some others are in the backlog of ingest work''.

Portico is not alone in this respect. As regards ``the backlog of ingest work'',
a study by the CLOCKSS Archive compared the dates at which content was
received at the archive from a major publisher via file transfer
with the data of publication. It concluded that approximately half the articles were received in the
year of publication, about a quarter in the year following publication,
and the remaining quarter over the decade following publication.

As regards ``lacking data needed to properly ingest them'',
the CLOCKSS Archive's experience suggests that the quality of
publisher-supplied metadata depends far more on the publisher
than on the mode of ingest.

The current file transfer content ingest pipeline for the CLOCKSS Archive is
similar in outline to those of other e-journal archives using file transfer.
It is documented in detail in the Archive's TRAC audit
materials~\cite{CLOCKSS-Audit}.
Briefly,
the process of ingesting journal content involves:
\begin{itemize}
\item Obtaining packages of content from each publisher by one of the
following techniques:
\begin{itemize}
\item \emph{Pull} FTP, in which the publisher gives the archive a
account on their FTP server.
The archive uses it to poll the FTP server to find and download
content newly placed there by the publisher.
\item \emph{Push} FTP, in which the archive gives the publisher an
account on an FTP server at the archive.
The publisher uses it to place content there,
which the archive subsequently ingests.
\item \emph{rsync}, in which rsync~\cite{rsync} is used to synchronize
a directory tree at the publisher with a directory tree at the archive.
The publisher adds content to their tree,
the archive ingests content that arrives in their tree.
\end{itemize}
\item At scale,
network transfers of content are not completely reliable.
Ideally,
the package format used by the publisher provides checksums
against which the received package can be validated.
Packages that fail validation should be discarded.
Some feedback mechanism is needed:
\begin{itemize}
\item In the Pull and rsync cases to tell the publisher when the
content has been successfully downloaded so that they can free up
the space it takes.
\item In the Push case to tell the publisher that the content they
sent was corrupted in transit and should be re-sent.
\end{itemize}
\item Some archives,
especially those which normalize the content for dissemination,
unpack the content from its publisher-specific format for storage.
Others,
especially dark archives,
postpone this step until the content is triggered for dissemination.
\item The metadata extractor~\cite{LOCKSS-MetadataExtractor}
is programmed to understand each publisher's package format,
and to know where in it to find the necessary metadata.
This may be in separate XML files or,
as with harvest content,
in the content itself.
It then creates an entry containing the metadata in a metadata
database~\cite{LOCKSS-MetadataDatabase}.
This may use heuristics to locate the metadata in one or more of the
article's files,
and does use heuristics
to normalize it into a canonical form~\cite{RosenthalMetadataIIPC2013}.
\end{itemize}

File transfer has the advantage that it uses fewer heuristics than
harvest,
but it has the disadvantage that it obtains content only when the
publisher deigns to send it,
rather than when it is made available to the readers.
In the real world,
both the heuristics and the publishers' processes are less than perfect.
It works well for the major publishers that use it,
but it is too complex for smaller publishers in the ``long tail''
of journals,
which are,
after all,
those at most risk and thus at most need of preservation.

Although,
per article,
the file transfer ingest process needs less monitoring and
adjustment than the harvest process,
it can't be left to operate unsupervised.
Automation of this monitoring would be greatly aided if there were
a second,
independent source of data against which the data extracted from
the content the publisher transferred could be compared.
To what extent can data obtained from CrossRef play this role?

\section{Building Blocks for a Solution}
\label{Sec:Solution}

The previous sections described the significant problems that e-journal archiving efforts face in the current environment, 
when trying to address the crucial Manifest, Completeness, and Bibliography (including Identity) requirements. These include:
\begin{itemize}
\item The lack of a reliable and machine-actionable start page to support collecting newly published objects.
\item The reliance on publisher-specific heuristics to determine links on human start pages that lead to newly published objects. 
These pages frequently link to many things other than new objects. 
\item The reliance on publisher-specific heuristics to determine the web boundary of new objects. 
The entry pages for these objects link to web resources that are part of the object but also to many resources that are not. 
In addition, different publishers have different implementations of these entry pages: sometimes they are landing pages providing an abstract, 
sometimes they are an HTML version of an article. 
\item The reliance on publisher-specific heuristics to obtain bibliographic metadata describing the new objects, including the object's identifier. 
Publisher metadata is of variable quality, can be located in a number of places (e.g., in HTML meta-tags, in the text, in dedicated 
bibliographic records arranged according to various metadata formats). 
\item In case of the File Transfer approach, the inability to verify whether content that is published is also made available (in a timely manner) 
by the publisher for archiving.  
\end{itemize}

This section outlines a solution to fundamentally address these problems. 
The solution involves new web infrastructure to be operated by CrossRef and academic publishers, and also 
leverages existing CrossRef infrastructure. And, although the focus of this paper is on e-journals, 
the proposed solution can also be implemented by institutional repositories, discipline repositories, 
and, more generally, portals that contribute to web-based scholarly communication. In addition, the solution 
has merits beyond e-journal archiving. It can empower other applications such as citation managers and alt metrics. 
And it can help DOIs achieve their full potential as persistent identifiers addressing the problem that, many times, 
DOI-identified objects are referenced by means of their locating URI instead of their persistent DOI at \msg{dx.doi.org} 
~\cite{vdSompel-WWW2016}. 

The proposed solution builds on two components:
\begin{itemize}
\item Machine-Actionable Change Feeds - As a means to provide information about new, changed, or deleted 
scholarly objects at an aggregate level, 
that is, at the level of CrossRef, a publisher's platform, a journal, an institutional repository, etc. 
\item Signposting - As a means to provide information about the resources that make up the web presence of 
a scholarly object, when HTTP-interacting with each of those resources individually. 
\end{itemize}

\subsection{Modeling}
\label{Sec:Solution:Modeling}

Both components crucially depend on an approach to model the various resources that make up the web presence of 
a scholarly object in a manner that can be applied across publishers, repositories, and portals; modeling resource types is addressed in 
Section \ref{Sec:Solution:Modeling:Resources}. Both components also crucially rely on typed links that can be used to 
interconnect these various web resources as a means to express 
identity of the object, its web boundary, the location of bibliographic metadata describing the object, etc.;   
typed links used for these purposes are addressed in Section \ref{Sec:Solution:Modeling:Relations}. 

\subsubsection{Resource Types and Resource Patterns}
\label{Sec:Solution:Modeling:Resources}

Several web resources of different types make up the distributed web presence of a scholarly digital object. 
For example, for most academic publications, a DOI at \msg{dx.doi.org} 
provides a persistent HTTP identity for the object, and a so-called landing 
page that describes the object serves as the page of first entry into a publisher's portal when dereferencing that DOI. 
But, sometimes the page of entry is not a landing page but rather an HTML version of a publication. Also, 
several institutional and discipline repositories do not assign DOIs but use local persistent HTTP URIs that are preferred for 
linking and citing. To address the Manifest, Completeness, and Bibliography requirements, 
these types of web resources must be distinguished: 

\begin{itemize}
\item \msg{Identifying HTTP URI} - An HTTP URI that provides a scholarly object with a web identity that is 
intended to be persistent. The persistent \msg{Identifying HTTP URI} redirects to a location URI 
that can change over time because the identified content moves from one web location to another, for example, 
because of platform migrations (location URI changes within the same domain) or transfer of content to a new custodian 
(location URI changes to another domain). There are two common approaches to achieve the intended persistence. One consists of 
operating shared, cross-domain, persistent identifier infrastructure. In this case, \msg{Identifying HTTP URIs} are minted in a dedicated domain 
and are redirected to locating URIs in a wide variety of domains. Examples include DOI, handle, identifiers.org, PURL, and W3ID. 
In this approach, the correspondence between the \msg{Identifying HTTP URI} and the locating URI is contained in a shared look-up table 
that is kept up-to-date by custodians of the identified content. Another approach consists of minting \msg{Identifying HTTP URIs} in the same domain as 
the locating URI. The correspondence between the \msg{Identifying HTTP URIs} and the locating URI is locally maintained, either in a look-up table 
or by means of web server rewrite rules. 
\item \msg{Entry Page} - The page in a publisher's portal where one typically enters when accessing a scholarly object, usually 
by following redirects from an \msg{Identifying HTTP URI} or via a search engine result. 
For many publishers, this \msg{Entry Page} is an HTML landing page that describes the object. But other publishers provide an 
HTML version of an article as the \msg{Entry Page}. 
\item \msg{Publication Resource} - A web resource provided by or via the publisher that is considered an integral part of the 
scholarly object. Examples include various renderings (HTML, PDF, PS, XML, etc.) of an article but also supplementary materials 
that are considered part of the object.
\item \msg{Bibliographic Resource} - A web resource that describes the object in a structured manner, for example, using 
BibTeX or RIS formatting styles. 
\end{itemize}

\begin{figure*}[htbp]
\centering{\includegraphics[width=1\textwidth,frame]{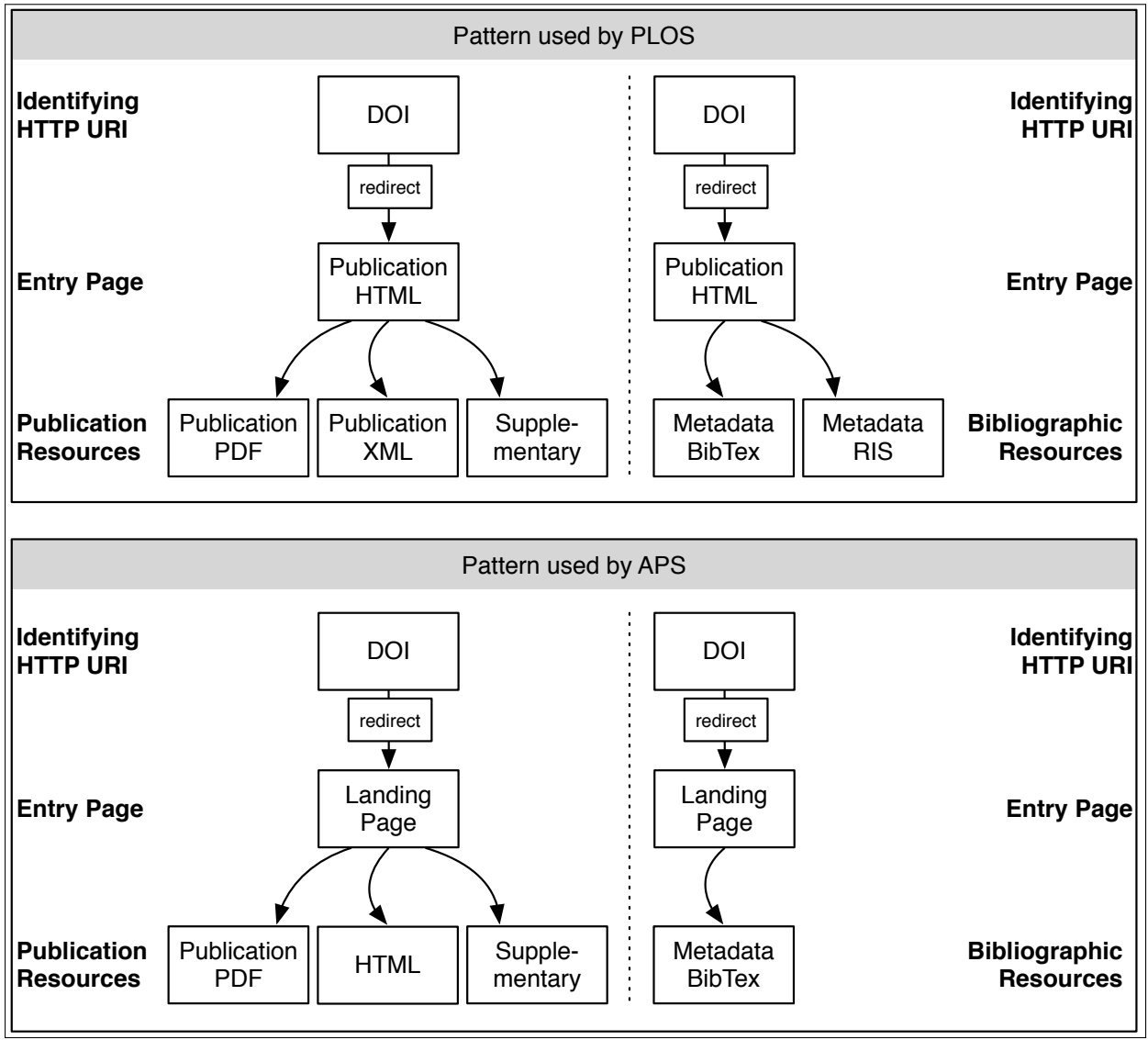}}
\caption{{\small Patterns used by PLOS and APS, respectively}}
\label{fig:Patterns1}
\end{figure*}

\begin{figure*}[htbp]
\centering{\includegraphics[width=1\textwidth,frame]{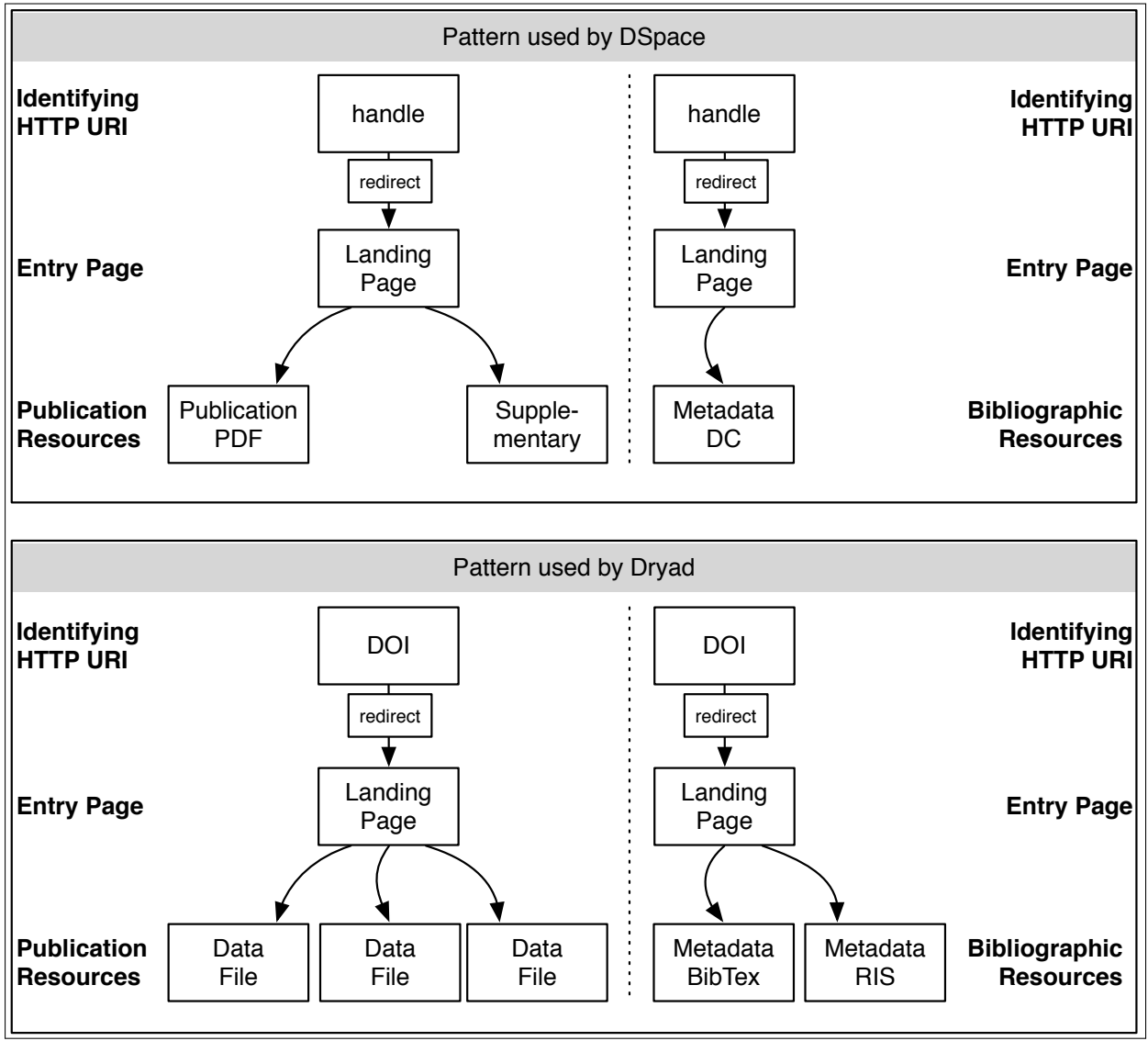}}
\caption{{\small Patterns used by DSpace and Dryad, respectively}}
\label{fig:Patterns2}
\end{figure*}

\begin{figure*}[htbp]
\centering{\includegraphics[width=1\textwidth,frame]{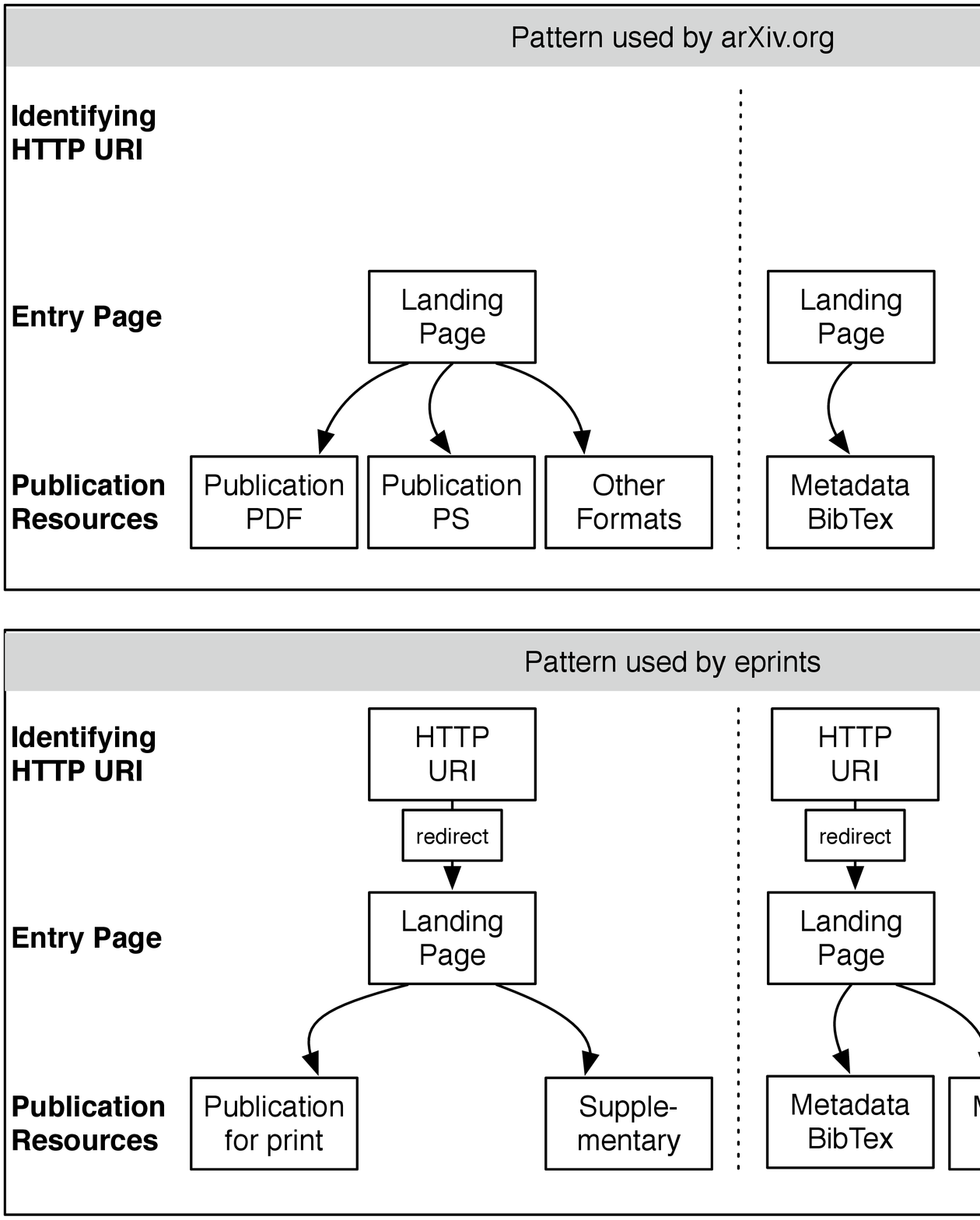}}
\caption{{\small Patterns used by arXiv.org and eprints, respectively}}
\label{fig:Patterns3}
\end{figure*}

Not all platforms have all of the above types of web resources associated with a digital object. 
Also, the way in which these resources are combined differs depending on the platform. Figures \ref{fig:Patterns1}, \ref{fig:Patterns2}, 
and \ref{fig:Patterns3} show various configurations as well as example platforms that use them. Each configuration shows the 
pattern involving \msg{Publication Resources} to the left, and the pattern with \msg{Bibliographic Resources} to the right. 

\subsubsection{Relation Types}
\label{Sec:Solution:Modeling:Relations}

In order to connect all resources that make up the web presence of a digital object, typed links are used. 
These links serve the purpose of expressing the type of a resource, when necessary, of indicating the identity of the 
object and delineating its web boundary, and to provide access to bibliographic metadata. Each typed link contains:
\begin{itemize}
\item The URI that is the source of the link. Sometimes, the source URI is 
provided implicitly such as, for example, in HTML where it is the URI of the document in 
which the link is embedded. Sometimes the source is provided explicitly. 
\item The URI that is the target of the link. A link between a source URI and a target URI
expresses that the target is somehow related to the source. 
\item The type of the relationship between the source and the target. 
This is expressed by means of the \msg{rel} attribute. Values for this attribute are strings registered in the 
IANA Link Relation Type Registry~\cite{IanaLinkRelationRegistry} and typically represent rather coarse relation types 
intended for web-wide interoperability. Values can also be URIs minted by communities, allowing for increased expressiveness  
but leading to decreased interoperability.
\item Optionally, additional attributes can be conveyed on the link, such as \msg{type} to express the \msg{mime-type} of the 
target resource and others listed in RFC5988~\cite{RFC5988}. 
\end{itemize}

Both proposed components - Change Feeds and Signposting - can and should use the same link relation types and link attributes. 
But the way in which these are conveyed differs per component:
\begin{itemize}
\item Machine-Actionable Change Feeds - The links and attributes are conveyed as part of a feed entry that is dedicated to 
a change event pertaining to a specific scholarly object.
\item Signposting - The links and attributes are conveyed in the HTTP Link header~\cite{RFC5988} when issuing an HTTP HEAD/GET on the resource.
\end{itemize}

Research conducted since the publication of ~\cite{vdSompel2015}, which introduced the Signposting concepts, 
has led to a shortlist of relation types - summarized in Table \ref{tab:reltypes} - 
that can be used to address the Manifest, Completeness, and Bibliography requirements. 
Links with these relation types must be applied to the various configurations discussed above 
to interlink all resources in such a way that a crawler or other application can unambiguously (and hence automatically) 
navigate its way around the web resources that make up a digital object. 
With one exception, all required relation types are already registered in the IANA Link Relation Type 
Registry~\cite{IanaLinkRelationRegistry}:

\begin{itemize}
\item \msg{item} \& \msg{collection}: In order to group all the publisher's \msg{Publication Resources} 
that are part of a scholarly object, the \msg{Entry Page} for the object is modeled as a collection 
that contains the other resources as items. This \msg{Entry Page} collection resource binds 
the other \msg{Publication Resources} by means of the \msg{item} relation type. 
These \msg{item} \msg{Publication Resources} in their turn link back to the \msg{Entry Page} 
collection resource using the \msg{collection} relation type. 
Which resources should be considered as items for an object is at a publisher's discretion but definitely included should be 
all renderings of an article (PDF, HTML, PS, ...) and any supplementary files considered to be an integral part of the object. 
\item \msg{persistent-id}: In order for the \msg{Entry Page} and the \msg{Publication Resources} 
to express that they are part of an object with an \msg{Identifying HTTP URI} they link to that 
 URI using the \msg{persistent-id} relation type. For example, for a DOI-identified object the target of this 
 \msg{persistent-id} is the DOI at \msg{dx.doi.org}. The \msg{persistent-id} relation type is not yet registered 
 but the plan is to author an RFC to define and register it.  At the time of writing, the name \msg{persistent-id} is used as a placeholder.
\item \msg{type}: In order for the \msg{Entry Page} and the \msg{Publication Resources} to express their own nature, 
the \msg{type} relation type is used. To avoid endless classification discussions, the proposal is to use as targets of 
these \msg{type} links URIs that are coarse indicators of the resource's own nature. 
The \msg{info:eu-repo/ vocabulary}\footnote{\url{https://wiki.surfnet.nl/display/standards/info-eu-repo}}, 
originated by SURF\footnote{\url{https://www.surf.nl/}} and meanwhile 
maintained by COAR\footnote{\url{https://www.coar-repositories.org/}} provides good basic candidates. For example, 
it uses the URI \msg{info:eu-repo/semantics/humanStartPage} (or a PURL variant) to indicate that a resource is a landing page, 
\msg{info:eu-repo/semantics/article} to convey it's an article, \msg{info:eu-repo/semantics/dataset} for a dataset,  
\msg{info:eu-repo/semantics/objectFile} for a file that is neither article nor dataset, and \msg{info:eu-repo/semantics/descriptiveMetadata} 
for bibliographic metadata. 
\item \msg{describedby} \& \msg{describes}: The \msg{describedby} relation type is used to link to a metadata resource 
that describes the scholarly object. A publisher uses it to link from the \msg{Entry Page} to a metadata resource 
and CrossRef uses it to link from a DOI at \msg{dx.doi.org} to its metadata that describes the DOI-identified object. 
The inverse relation type \msg{describes} is used to link back from the metadata resource. Following the same examples, 
this links from a publisher's metadata record to the \msg{Entry Page} and from CrossRef's metadata to the 
DOI at \msg{dx.doi.org}. 
\end{itemize}

For typed links that use these relation types, the addition of the following attributes - summarized in Table \ref{tab:reltypes} - 
can be important. The first attribute is already widely used, and work is under way to 
define the two additional attributes\footnote{\url{https://github.com/mnot/I-D/issues/175}}~\cite{link-hints}.

\begin{itemize}
\item \msg{type}: In order to express the \msg{mime-type} of the target resource, the \msg{type} attribute is used. 
This can, for example, be very useful on \msg{item} links pointing at \msg{Publication Resources} to allow 
distinguishing between, e.g., the PDF, HTML, PS version of an article.
\item \msg{profile}: The \msg{profile} attribute can be used to provide further details about the representation of a 
linked resource. It can be especially helpful for \msg{Bibliographic Resources}. Many 
have \msg{text/plain}, \msg{application/xml}, and, more recently, \msg{application/json} as \msg{mime-type} 
and it would be helpful to further inform clients about the specific nature of a format, 
for example \msg{BibTeX}, \msg{RIS}, \msg{Dublin Core}, or \msg{MODS}. 
This can be achieved by adding a \msg{profile} attribute to the \msg{describedby} links with as value a URI 
that uniquely identifies the format, e.g., \url{http://www.bibtex.org/} for \msg{BibTeX}, and an XML Schema URI 
for XML metadata.
\item \msg{sem-type} As mentioned above, a link with a \msg{type} relation type is used to 
point from a resource to a URI that expresses the resource's own nature, e.g., whether it is a landing page, an article, etc. 
The \msg{sem-type} attribute can be used on a link to express the nature of the target resource. As such, while the \msg{type} 
attribute on a link expresses the \msg{mime-type} of the linked resource (e.g., \msg{application/pdf}), 
the \msg{sem-type} attribute expresses its semantic type (e.g., \msg{info:eu-repo/semantics/article}). 
At the time of writing, the name \msg{sem-type} is used as a placeholder.
\end{itemize}

\begin{table}
\footnotesize
\centering
\begin{tabular}{|l|l|}\hline
\textbf{Link Relation Type} & \textbf{Attributes} \\\hline
\msg{collection} & \msg{type}, \msg{profile}, \msg{sem-type} \\\hline
\msg{describedby} & \msg{type}, \msg{profile}, \msg{sem-type} \\\hline
\msg{describes} & \msg{type}, \msg{profile}, \msg{sem-type} \\\hline
\msg{item} & \msg{type}, \msg{profile}, \msg{sem-type} \\\hline
\msg{persistent-id} &  \\\hline
\msg{type} & \msg{type}, \msg{profile}, \msg{sem-type} \\\hline
\end{tabular}
\caption{\label{tab:reltypes}Relation types and attributes to address the Manifest, Completeness, and Bibliography requirements.}
\end{table}

\begin{figure*}[htbp]
\centering{\includegraphics[width=1\textwidth,frame]{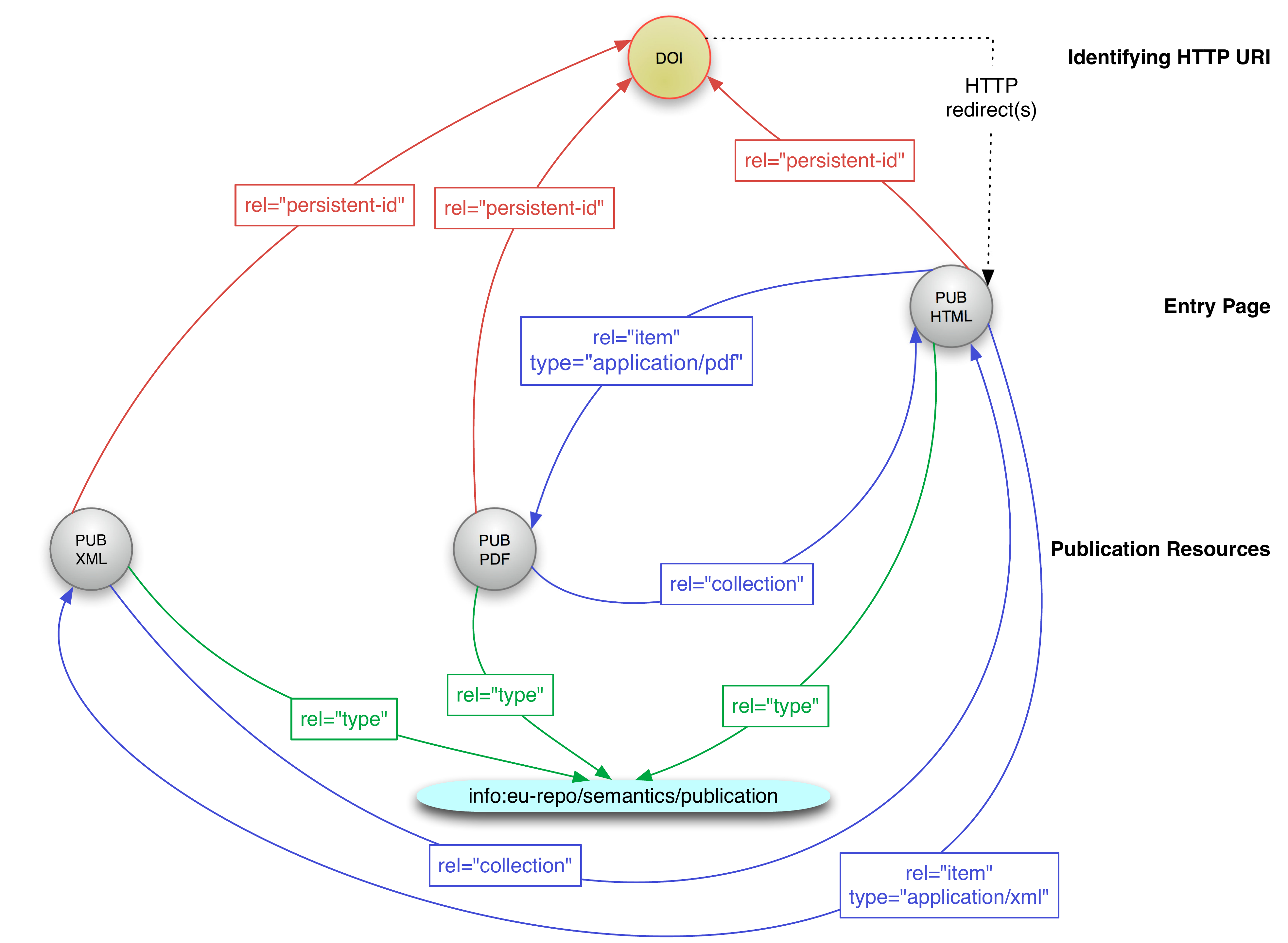}}
\caption{{\small Typed links to connect resources that make up the web presence of a scholarly digital object.}}
\label{Fig:PLoS_Style}
\end{figure*}

\begin{figure*}[htbp]
\centering{\includegraphics[width=1\textwidth, frame]{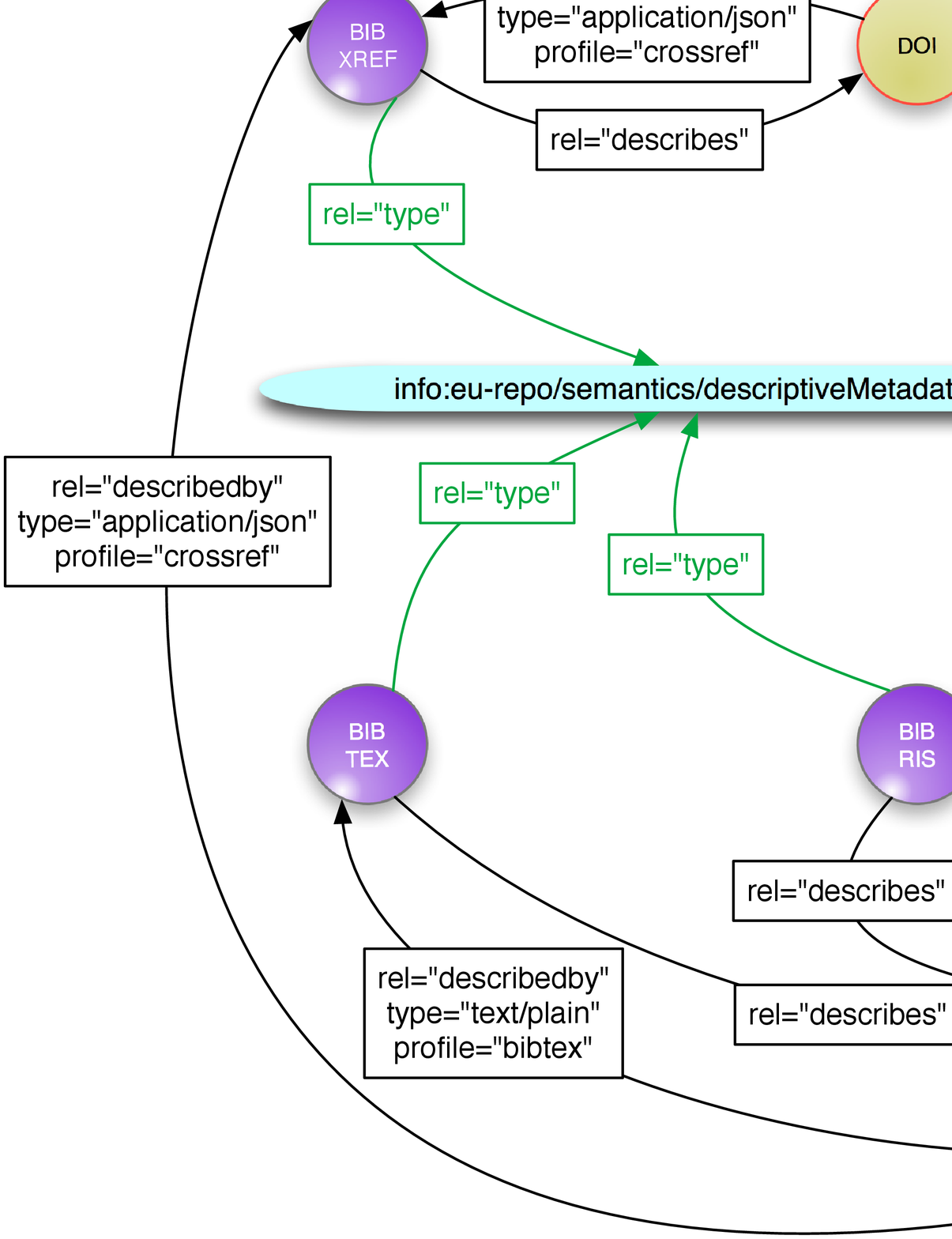}}
\caption{{\small Typed links to support discovery of bibliographic metadata.}}
\label{Fig:Biblio_PLoS_Style}
\end{figure*}

Typical relationships among web resources that make up a scholarly article 
are shown in Figure~\ref{Fig:PLoS_Style}. The figure depicts a pattern 
used by several publishers, including PLOS. Relations pertaining to 
identification (\msg{persistent-id}) are in red. 
Links pertaining to delineating the web boundary of an object (\msg{item} \& \msg{collection}) are shown in blue, 
and the ones expressing the nature of a resource ({\msg{type}) in green. In this pattern, the \msg{Entry Page} 
is the HTML version of the article. A crawler that has followed redirects from the DOI at \msg{dx.doi.org} and has arrived at the \msg{Entry Page} 
can use links in that page's HTTP header to find all \msg{Publication Resources} without having to screen-scrape the page. 
In this pattern, \msg{Publication Resources} can be gathered by following the \msg{item} links and by collecting the 
\msg{Entry Page} itself because it has a \msg{type} link that identifies itself as an article. 
Note also how the \msg{collection} links allow a crawler to collect all \msg{Publication Resources} 
even if it did not land on the \msg{Entry Page}. Similarly, the \msg{persistent-id} links 
allow the crawler to determine the \msg{Identifying HTTP URI} of a \msg{Publication Resource} 
when it lands on one. These links are also very useful for 
applications other than archival crawlers. For example, they allow a citation manager to automatically detect the DOI 
of an article, even when landing on its PDF version. 

The Figure also shows the use of the \msg{type} attribute 
on \msg{item} links to convey the \msg{mime-type} of a linked resource. This can significantly assist selective crawlers,
which only want to collect components of the object suitable for their processes. 
An example would be a crawler supporting data-mining, which only needs the text and would much prefer HTML over 
PDF~\cite{Brook-11-2014}. The \msg{sem-type} attribute, not shown in the figure, would support similar selectivity regarding 
the nature of a linked resource rather than its \msg{mime-type}. 

Figure~\ref{Fig:Biblio_PLoS_Style} shows the relationships pertaining to \msg{Bibliographic Resources} 
for the same pattern. It shows two metadata records exposed by the publisher. Note that it is essential that a 
metadata record be accessible at a URI by using HTTP GET. Strangely enough, some publishers use a form-based  
 HTTP POST approach for metadata access. 
For brevity, the figure merely shows strings (e.g., \msg{BibTeX}) as the value of the \msg{profile} attribute, in reality, 
a URI (e.g., \url{http://www.bibtex.org/}) must be used. Note also that, in the figure, CrossRef provides a \msg{describedby} link 
pointing from the DOI at \msg{dx.doi.org} at its metadata that describes a DOI-identified object. In addition, 
the publisher also links to that metadata. It can do so because the URI of CrossRef's metadata can be constructed 
when the DOI is known. All these links support discovery of metadata without out-of-band knowledge. 

\subsection{Web Infrastructure}
\label{Sec:Solution:Infrastructure}

This section introduces concrete technologies that can be used to address the 
Manifest, Completeness, and Bibliography requirements: the CrossRef API~\cite{CrossRefApi}, 
ANSI/NISO Z39.99 ResourceSync~\cite{ResourceSync-spec}, and 
Signposting. These technologies are used in a manner that leverages the modeling approach described above. 
The section also highlights which technology addresses which of the three requirements; this is summarized in Table \ref{tab:infra}.

\begin{table}
\footnotesize
\begin{tabular}{|l|c|c|c|}\hline
\textbf{Infrastructure}  & \textbf{Manifest} & \textbf{Completeness} & \textbf{Bibliography} \\\hline
           & \multicolumn{3}{c|}{\textbf{Implemented by CrossRef}} \\\hline
CrossRef API & +\textbackslash- & - & + \\\hline
ResourceSync & + & - & + \\\hline
Signposting & - & - & + \\\hline
           & \multicolumn{3}{c|}{\textbf{Implemented by Publisher}} \\\hline
CrossRef API & - & - & - \\\hline
ResourceSync & + & + & + \\\hline
Signposting & - & + & + \\\hline
\end{tabular}
\caption{\label{tab:infra}Infrastructure to address the Manifest, Completeness, and Bibliography requirements}
\end{table}

\subsubsection{CrossRef API}
\label{Sec:Solution:Infrastructure:API}

The CrossRef API provides two functionalities that can be used as building blocks for 
a solution:
\begin{itemize}
\item A means to retrieve recently registered DOIs and associated metadata, which can address the Manifest and Bibliography 
requirements.
\item A means to access metadata about a DOI-identified object, which can address the Bibliography requirement.
\end{itemize}

Listing \ref{lst:CrossRefAPI} shows an API query issued on March 17, 2016 
that returned the then 20 most recently deposited DOIs; only two entries are shown. 

\begin{lstlisting}[caption={CrossRef API used to retrieve the 20 most recently registered DOIs}, label=lst:CrossRefAPI, float=*, frame=single, captionpos=t]
~\textbf{Request}~

wget -O query.out http://api.crossref.org/works?sort=deposited&order=desc

~\textbf{Response}~

{ "message" : { "facets" : {  },
      "items" : [ { "DOI" : "10.1029/jd094id06p08425",
            "ISSN" : [ "0148-0227" ],
            "URL" : "http://dx.doi.org/10.1029/jd094id06p08425",
            "archive" : [ "Portico" ],
            "author" : [ { "affiliation" : [  ],
                  "family" : "Livingston",
                  "given" : "John M."
                },
                { "affiliation" : [  ],
                  "family" : "Russell",
                  "given" : "Philip B."
                }
              ],
            "container-title" : [ "Journal of Geophysical Research: Atmospheres",
                "J. Geophys. Res."
              ],
            "created" : { "date-parts" : [ [ 2008,
                      2,
                      6
                    ] ],
                "date-time" : "2008-02-06T13:40:44Z",
                "timestamp" : 1202305244000
              },
            "deposited" : { "date-parts" : [ [ 2016,
                      3,
                      17
                    ] ],
                "date-time" : "2016-03-17T19:59:58Z",
                "timestamp" : 1458244798000
              },
            "indexed" : { "date-parts" : [ [ 2016,
                      3,
                      17
                    ] ],
                "date-time" : "2016-03-17T20:40:08Z",
                "timestamp" : 1458247208393
              },
            "issue" : "D6",
            "issued" : { "date-parts" : [ [ 1989,
                      6,
                      20
                    ] ] },
            "license" : [ { "URL" : "http://doi.wiley.com/10.1002/tdm_license_1",
                  "content-version" : "tdm",
                  "delay-in-days" : 0,
                  "start" : { "date-parts" : [ [ 1989,
                            6,
                            20
                          ] ],
                      "date-time" : "1989-06-20T00:00:00Z",
                      "timestamp" : 614304000000
                    }
                },
                { "URL" : "http://onlinelibrary.wiley.com/termsAndConditions",
                  "content-version" : "vor",
                  "delay-in-days" : 8494,
                  "start" : { "date-parts" : [ [ 2012,
                            9,
                            21
                          ] ],
                      "date-time" : "2012-09-21T00:00:00Z",
                      "timestamp" : 1348185600000
                    }
                }
              ],
            "link" : [ { "URL" : "http://api.wiley.com/onlinelibrary/tdm/v1/articles/10.1029%2FJD094iD06p08425",
                  "content-type" : "application/pdf",
                  "content-version" : "vor",
                  "intended-application" : "text-mining"
                } ],
            "member" : "http://id.crossref.org/member/311",
            "page" : "8425-8433",
            "prefix" : "http://id.crossref.org/prefix/10.1002",
            "published-online" : { "date-parts" : [ [ 2012,
                      9,
                      21
                    ] ] },
            "published-print" : { "date-parts" : [ [ 1989,
                      6,
                      20
                    ] ] },
            "publisher" : "Wiley-Blackwell",
            "reference-count" : 16,
            "score" : 1.0,
            "source" : "CrossRef",
            "subject" : [ "Earth and Planetary Sciences (miscellaneous)",
                "Space and Planetary Science",
                "Atmospheric Science",
                "Geophysics"
              ],
            "subtitle" : [  ],
            "title" : [ "Retrieval of aerosol size distribution moments from multiwavelength particulate extinction measurements" ],
            "type" : "journal-article",
            "volume" : "94"
          },
          ...
\end{lstlisting}

\begin{lstlisting}[float=*, frame=single]
          { "DOI" : "10.1016/j.jmaa.2016.03.023",
            "ISSN" : [ "0022-247X" ],
            "URL" : "http://dx.doi.org/10.1016/j.jmaa.2016.03.023",
            "alternative-id" : [ "S0022247X16002419" ],
            "author" : [ { "affiliation" : [  ],
                  "family" : "Li",
                  "given" : "C."
                },
                { "affiliation" : [  ],
                  "family" : "Ng",
                  "given" : "K.F."
                }
              ],
            "container-title" : [ "Journal of Mathematical Analysis and Applications" ],
            "created" : { "date-parts" : [ [ 2016,
                      3,
                      17
                    ] ],
                "date-time" : "2016-03-17T19:58:50Z",
                "timestamp" : 1458244730000
              },
            "deposited" : { "date-parts" : [ [ 2016,
                      3,
                      17
                    ] ],
                "date-time" : "2016-03-17T19:58:50Z",
                "timestamp" : 1458244730000
              },
            "indexed" : { "date-parts" : [ [ 2016,
                      3,
                      17
                    ] ],
                "date-time" : "2016-03-17T20:41:11Z",
                "timestamp" : 1458247271154
              },
            "issued" : { "date-parts" : [ [ 2016,
                      3
                    ] ] },
            "license" : [ { "URL" : "http://www.elsevier.com/tdm/userlicense/1.0/",
                  "content-version" : "tdm",
                  "delay-in-days" : 0,
                  "start" : { "date-parts" : [ [ 2016,
                            3,
                            1
                          ] ],
                      "date-time" : "2016-03-01T00:00:00Z",
                      "timestamp" : 1456790400000
                    }
                } ],
            "link" : [ { "URL" : "http://api.elsevier.com/content/article/PII:S0022247X16002419?httpAccept=text/xml",
                  "content-type" : "text/xml",
                  "content-version" : "vor",
                  "intended-application" : "text-mining"
                },
                { "URL" : "http://api.elsevier.com/content/article/PII:S0022247X16002419?httpAccept=text/plain",
                  "content-type" : "text/plain",
                  "content-version" : "vor",
                  "intended-application" : "text-mining"
                }
              ],
            "member" : "http://id.crossref.org/member/78",
            "prefix" : "http://id.crossref.org/prefix/10.1016",
            "published-print" : { "date-parts" : [ [ 2016,
                      3
                    ] ] },
            "publisher" : "Elsevier BV",
            "reference-count" : 35,
            "score" : 1.0,
            "source" : "CrossRef",
            "subject" : [ "Applied Mathematics",
                "Analysis"
              ],
            "subtitle" : [  ],
            "title" : [ "Extended Newton methods for conic inequalities: Approximate solutions and the extended Smale \alpha-theory" ],
            "type" : "journal-article"
          },
          ...
  "message-type" : "work-list",
  "message-version" : "1.0.0",
  "status" : "ok"
}
\end{lstlisting}

This output illustrates some of the problems of using the CrossRef API to address the Manifest requirement. 
The second entry \msg{doi:10.1016/j.jmaa.2016.03.023} is a current paper
published by Elsevier, which an archive would want to ingest.
The first entry \msg{doi:10.1029/jd094id06p08425} was published June 20,
1989. The 10.1029 prefix belongs to the American Geophysical Union,
whose member number is 13. The member number 311 is Wiley-Blackwell.
This entry is presumably the result of a journal transfer,
about which archives need to be informed but which may or may not
require that the journal's back content be re-ingested,
depending upon arrangements with the previous and new publisher. 
As such, the fact that the API response does not discriminate between newly registered and
changed DOIs is problematic. 

Archives could poll this API, concatenate and deduplicate the results,
filter for the publishers they preserve, and synthesize what they really want,
which is a feed of newly registered DOIs for their publishers.
Using this API query would also address the Bibliography requirement. 
But, since this approach to address the Manifest requirement involves recurrent queries 
against the entire CrossRef database, it may not be the most efficient approach for either CrossRef or the
users of their API. The next section describes how the use of the ANSI/NISO Z39.99 ResourceSync 
standard is an appealing alternative. 

Listing \ref{lst:CrossRefAPImeta} shows how the CrossRef API is used to return JSON bibliographic
metadata pertaining to a DOI-identified object. Assuming the DOI of a newly registered or changed DOI is known, 
this API query can be used to address the Bibliography requirement. 
This metadata is obtained without using heuristics and CrossRef applies validation processes to it on receipt. 
This metadata can be used as is but can also be compared to metadata that might be exposed by publishers 
as \msg{Bibliographic Resources}. 

\begin{lstlisting}[caption={CrossRef API used to retrieve metadata about a DOI-identified article}, label={lst:CrossRefAPImeta}, float=*, frame=single, captionpos=b, 
basicstyle=\ttfamily\scriptsize]

~\textbf{Request}~

wget http://api.crossref.org/works/10.1045/september2015-rosenthal

~\textbf{Response}~

{ "message" : { "DOI" : "10.1045/september2015-rosenthal",
      "ISSN" : [ "1082-9873" ],
      "URL" : "http://dx.doi.org/10.1045/september2015-rosenthal",
      "author" : [ { "affiliation" : [  ],
            "family" : "Rosenthal",
            "given" : "David S. H."
          },
          { "affiliation" : [  ],
            "family" : "Vargas",
            "given" : "Daniel L."
          },
          { "affiliation" : [  ],
            "family" : "Lipkis",
            "given" : "Tom A."
          },
          { "affiliation" : [  ],
            "family" : "Griffin",
            "given" : "Claire T."
          }
        ],
      "container-title" : [ "D-Lib Magazine" ],
      "created" : { "date-parts" : [ [ 2015,
                9,
                15
              ] ],
          "date-time" : "2015-09-15T11:09:53Z",
          "timestamp" : 1442315393000
        },
      "deposited" : { "date-parts" : [ [ 2015,
                9,
                15
              ] ],
          "date-time" : "2015-09-15T12:46:38Z",
          "timestamp" : 1442321198000
        },
      "indexed" : { "date-parts" : [ [ 2015,
                12,
                22
              ] ],
          "date-time" : "2015-12-22T03:16:21Z",
          "timestamp" : 1450754181783
        },
      "issue" : "9/10",
      "issued" : { "date-parts" : [ [ 2015,
                9
              ] ] },
      "member" : "http://id.crossref.org/member/72",
      "prefix" : "http://id.crossref.org/prefix/10.1045",
      "published-online" : { "date-parts" : [ [ 2015,
                9
              ] ] },
      "publisher" : "CNRI Acct",
      "reference-count" : 0,
      "score" : 1.0,
      "source" : "CrossRef",
      "subject" : [ "Library and Information Sciences" ],
      "subtitle" : [  ],
      "title" : [ "Enhancing the LOCKSS Digital Preservation Technology" ],
      "type" : "journal-article",
      "volume" : "21"
    },
  "message-type" : "work",
  "message-version" : "1.0.0",
  "status" : "ok"
}
\end{lstlisting}

\subsubsection{ResourceSync}
\label{Sec:Solution:Infrastructure:ResourceSync}

ANSI/NISO Z39.99 ResourceSync \cite{ResourceSync-spec} was published in 2014, as the result of a 3-year joint effort by NISO and 
the Open Archives Initiative aimed at devising a ``webby'' successor to OAI-PMH~\cite{OAI-PMH}, the popular Protocol for Metadata Harvesting. 
ResourceSync is about synchronizing web resources - anything with a URI - across systems, 
not merely about synchronizing repository metadata like its predecessor OAI-PMH. 
But, since metadata can and is made available at URIs, ResourceSync can also be used to synchronize metadata. 
ResourceSync is based on the Sitemap protocol \cite{sitemap} that is used by search engines to index the web, 
but extends that protocol in three ways that make it rather attractive for the purpose of the e-journal archiving use case:

\begin{itemize}
\item The Sitemap protocol only allows for the publication of an inventory of web resources, essentially a list of URIs currently exposed by a system. 
This basic capability is called Resource Lists in ResourceSync. But, in addition, ResourceSync also allows 
the publication of Change Lists (lists of recently created, updated, or deleted resources), and Change Notifications 
(resource change events communicated on a PubSubHubbub channel to which applications subscribe).  
While the payloads in these communications point at resources by means of their URI, 
ResourceSync also provides Dump approaches that allow  packaging resource representations in ZIP files while conveying their web URIs 
in a manifest embedded in the ZIP. For example, a Change Dump points at a ZIP file of recently created/updated/deleted resources. 
Change Lists, Change Notifications, and Change Dumps are standard-based approaches to implement Machine-Actionable Change Feeds. 
All can play a role for crawl-based as well as file transfer archiving.
\item ResourceSync defines a variety of elements and attributes in support of resource synchronization, including ways to express 
fixity information, to detail the temporal interval covered by a Change List, to convey the datetime and nature 
(created, updated, deleted) of a change. 
\item ResourceSync allows to express typed links from a resource subject to 
synchronization to related resources, for example, a link from an article to metadata describing it. As such, all the 
relation types (and attributes) described in the above can be used in ResourceSync payloads.
\end{itemize}

CrossRef can use ResourceSync Change Lists to address the Manifest and Bibliography 
requirements by communicating about newly registred, changed, and deleted (if such a thing exists) DOIs. 
This can be done by pre-computation instead of by evaluating a query against its database at each poll. 
The use of ResourceSync Change Notifications would additionally eliminate polling.  
Listing \ref{lst:CrossRefPLOS} in Section \ref{Sec:Summary} illustrates how CrossRef can use ResourceSync to report 
on newly registered DOIs using ResourceSync.

Publishers can use ResourceSync Change Lists or Change Notifications 
to communicate about newly published objects. 
These ResourceSync feeds would address the Manifest requirement by serving as a machine-actionable, article-centric, 
alternative to the current journal ``start page'' visited 
by the archival web crawler. But, since ResourceSync can provide typed links to related resources, it can 
also be used to address the Completeness and Bibliography requirements. Moreover, since ResourceSync 
also supports reporting on changed and deleted resources, publishers could use it to generally keep 
applications informed about their evolving content. 
In addition, since ResourceSync allows expressing fixity information, 
it could also support verification of collected resources. 
Listing \ref{lst:PLOSevent} in Section \ref{Sec:Summary} shows how a publisher can 
use ResourceSync to announce the availability of a new article. 

If CrossRef, publishers, institutional and discipline repositories, as well as other portals involved 
in web-based scholarly communication would uniformly 
use ResourceSync to report on holdings and changes to holdings,
clients would have a uniform method to stay informed about evolving scholarly content. 
This was the original motivation for creating the ResourceSync standard in the first place. 
But, unfortunately, so far, ResourceSync has not enjoyed wide adoption despite its power and simplicity. 
Possible explanations include competition with its significantly less powerful and 15 year old predecessor OAI-PMH, 
an overall tendency by portals to resort to rolling their own bespoke APIs, 
and a lack of concrete action regarding interoperability despite a general sentiment that more is needed. 

\subsubsection{Signposting}
\label{Sec:Solution:Infrastructure:Signposting}

A recent paper by Van de Sompel and Nelson~\cite{vdSompel2015} shows how simple
changes to the publishing platforms can address the Completeness and Bibliography requirements. 
Their ``Signposting the Scholarly Web'' proposal uses links conveyed in 
HTTP Link headers, as defined in RFC5988~\cite{RFC5988}, 
to interconnect the web resources that make up a scholarly digital object.
The essence of the Signposting idea is that each such web resource 
can provide informative typed links in an HTTP Link header in responses to HTTP HEAD/GET requests. 

In order to address the Completeness requirement, publishers can express typed links and attributes, 
as described in Section~\ref{Sec:Solution:Modeling:Relations} and as illustrated in Figure~\ref{Fig:PLoS_Style},  
using HTTP Link headers when a client issues an HTTP HEAD/GET on 
the \msg{Entry Page} and \msg{Publication Resources}. These links 
allow a crawler or other application to unambiguously navigate across the web resources that make up a scholarly object. 
Listing \ref{lst:PLOSmetalink} in Section \ref{Sec:Summary} shows a response header for the \msg{Entry Page}, 
which includes links to the \msg{Identifying HTTP URI} and to each of the 
\msg{Publication Resources}. Another link indicates that the \msg{Entry Page} is actually an article. 
Listing \ref{lst:PLOSpdf} shows a response header for   
the PDF version of an article containing links that indicate the PDF is actually an article, 
what its \msg{Identifying HTTP URI} is, and where the \msg{Entry Page} can be found from 
which other resources that are part of the object can be discovered.

In order to address the Bibliography requirement, CrossRef can include a link in the HTTP response header for a 
DOI at \msg{dx.doi.org} that points with the \msg{describedby} relation type to the URI of the metadata record. 
This link should also have the \msg{type} attribute to convey the \msg{mime-type} of 
CrossRef's metadata, i.e., \msg{application/json}, and the \msg{profile} attribute to provide 
further information about the specific JSON used by CrossRef. Listings \ref{lst:CrossRefmetalink} 
in Section \ref{Sec:Summary} provides an illustration. 
The metadata record can link back to the DOI at \msg{dx.doi.org} using the inverse \msg{describedby} relation type.  
Publishers can add exactly the same \msg{describedby} link to point from an  
\msg{Entry Page} to the guessable URI for metadata at CrossRef's API. 
In addition, bibliographic metadata exposed by publishers can 
be handled in a similar way by including \msg{describedby} links, each pointing from the \msg{Entry Page} to the 
URI of a metadata record expressed according to a specific format. Again, these links should use the \msg{type} and 
\msg{profile} attributes. Listing \ref{lst:PLOSmetalink} in Section \ref{Sec:Summary} illustrates these links. 
 The URIs of bibliographic records both at CrossRef and at the publisher can further include a link in their HTTP response header 
that characterizes the resources as bibliographic metadata (\msg{info:eu-repo/semantics/descriptiveMetadata} 
in the info:eu-repo vocabulary) using the \msg{type} relation type. 

\section{Summary}
\label{Sec:Summary}

The remainder of this section provides a concrete insight into how CrossRef and publishers can 
support the Manifest, Completeness, and Bibliography requirements by implementing ResourceSync and 
Signposting, and by leveraging existing CrossRef API functionality. The examples shown below are for the PLOS One paper 
``Scholarly Context Not Found: One in Five Articles Suffers from Reference Rot'' \cite{EvanescentWeb}
for which Table \ref{tab:PLOSURIS} shows the URIs of web resources that make up the paper's web presence. 
The examples use all the relation types and attributes discussed above.

\begin{table*}
\footnotesize
\begin{tabular}{|l|p{0.8\linewidth}|}\hline
\textbf{Resource}  & \textbf{Resource URI} \\\hline
\multicolumn{2}{|c|}{\textbf{CrossRef}} \\\hline
DOI  & \url{http://dx.doi.org/10.1371/journal.pone.0115253} \\\hline
CrossRef metadata & \url{http://api.crossref.org/works/10.1371/journal.pone.0115253} \\\hline
\multicolumn{2}{|c|}{\textbf{Publisher - PLOS}} \\\hline
HTML article & \url{http://journals.plos.org/plosone/article?id=10.1371/journal.pone.0115253} \\\hline
PDF article & \url{http://journals.plos.org/plosone/article/asset?id=10.1371%2Fjournal.pone.0115253.PDF} \\\hline
XML article & \url{http://journals.plos.org/plosone/article/asset?id=10.1371%2Fjournal.pone.0115253.XML} \\\hline
Supplementary file & \url{http://journals.plos.org/plosone/article/asset?unique&id=info:doi/10.1371/journal.pone.0115253.s012} \\\hline
PLOS BibTeX metadata & \url{http://journals.plos.org/plosone/article/citation/bibtex?id=10.1371%2Fjournal.pone.0115253} \\\hline
PLOS RIS metadata & {\url{http://journals.plos.org/plosone/article/citation/ris?id=10.1371%2Fjournal.pone.0115253} }\\\hline
\end{tabular}
\caption{\label{tab:PLOSURIS}URIs of resources that make up the web presence of the PLOS One example article \cite{EvanescentWeb}}
\end{table*}

\subsection{ResourceSync}
\label{Sec:Summary:Machine-Actionable-Feeds}

Presume CrossRef uses ResourceSync Change Lists or Change Notifications to convey newly registered DOIs. 
The payload to convey the new PLOS One paper is shown in Listing \ref{lst:CrossRefPLOS}. Note that 
the design choice is made to express the payload in terms of the \msg{Identifying HTTP URI} (the DOI at \msg{dx.doi.org} in 
the \msg{loc} element) and include a link from that URI to the metadata record that describes the DOI-identified object.
As such, ResourceSync is used to communicate about change events in the DOI namespace. These can include 
new DOI registration (\msg{created} event), changes to a registration (\msg{updated} event), and deletions of a DOI (\msg{deleted} event), 
if such a  thing exists. 

\begin{lstlisting}[label={lst:CrossRefPLOS}, caption={Payload for a ResourceSync change event in which CrossRef announces a new DOI}, frame=single, float=*,
basicstyle=\ttfamily\scriptsize]
<url>
  <loc>http://dx.doi.org/10.1371/journal.pone.0115253</loc>
  <rs:md change="created" datetime="2014-12-26T00:00:00Z"/>
  <!-- link to metadata describing the DOI -->
  <rs:ln rel="describedby" 
         href="http://api.crossref.org/works/10.1371/journal.pone.0115253" 
         type="application/json" 
         profile="https://github.com/CrossRef/rest-api-doc"/>
</url>
\end{lstlisting}

The publisher can also use ResourceSync Change Lists or Change Notifications to convey the publication of a new paper. 
Listing \ref{lst:PLOSevent} shows an intentionally rich payload that does this for paper \cite{EvanescentWeb}, 
and that is expressed in terms of the URI of the \msg{Entry Page} (in this case the HTML article) modelled as a collection. 
Links have the URI of the \msg{Entry Page} (the URI in the \msg{loc} element) as source and express the nature of the 
\msg{Entry Page} (\msg{info:eu-repo/semantics/article}), reveal the \msg{Identifying HTTP URI}, and provide a pathway to \msg{Publication Resources} and 
\msg{Bibliographic Resources}. Note that the \msg{mime-type} of linked resources is conveyed using the 
\msg{type} attribute, and their nature using the \msg{sem-type} attribute.

In addition to the information shown, ResourceSync also allows expressing fixity information pertaining to 
the resource that is the subject of the payload (the URI in the \msg{loc} element) as well as pertaining to linked resources. 
Changes to the paper, for example, an update or deletion of a \msg{Publication Resources} can be communicated in a similar way. 
In this case, however, the change event should be expressed in terms of the URI of the changed resource (the URI in the \msg{loc} element), 
and links to neighboring resources (\msg{Entry Page}, \msg{Persistent HTTP URI}) should be provided.

\begin{lstlisting}[label={lst:PLOSevent}, caption={Payload for a ResourceSync change event in which PLOS announces a new paper}, frame=single, float=*, 
basicstyle=\ttfamily\scriptsize]
<url>
  <loc>http://journals.plos.org/plosone/article?id=10.1371/journal.pone.0115253</loc>
  <rs:md change="created" datetime="2014-12-26T00:00:00Z"/>
  <!-- link that categorizes the resource with the URI in loc as an article -->
    <rs:ln rel="type" 
         href="info:eu-repo/semantics/article" />
  <!-- link to the DOI -->
  <rs:ln rel="persistent-id" 
         href="http://dx.doi.org/10.1371/journal.pone.0115253" />
  <!-- link to the PDF article -->
  <rs:ln rel="item" 
         href="http://journals.plos.org/plosone/article/asset?id=10.1371%2Fjournal.pone.0115253.PDF" 
         type="application/pdf" 
         sem-type="info:eu-repo/semantics/article"/>
  <!-- link to the XML article -->
  <rs:ln rel="item" 
         href="http://journals.plos.org/plosone/article/asset?id=10.1371%2Fjournal.pone.0115253.XML" 
         type="application/xml" 
         sem-type="info:eu-repo/semantics/article"/>
  <!-- link to the supplementary file -->
  <rs:ln rel="item" 
         href="http://journals.plos.org/plosone/article/asset?unique&id=info:doi/10.1371/journal.pone.0115253.s012" 
         type="text/html" 
         sem-type="info:eu-repo/semantics/objectFile"/>
  <!-- link to CrossRef metadata -->
  <rs:ln rel="describedby" 
         href="http://api.crossref.org/works/10.1371/journal.pone.0115253"
         type="application/json"
         profile="https://github.com/CrossRef/rest-api-doc"/>
  <!-- link to PLOS BibTeX metadata -->
  <rs:ln rel="describedby" 
         href="http://journals.plos.org/plosone/article/citation/bibtex?id=10.1371%2Fjournal.pone.0115253"
         type="text/plain"
         profile="http://bibtex.org"/>
  <!-- link to PLOS RIS metadata -->
  <rs:ln rel="describedby" 
         href="http://journals.plos.org/plosone/article/citation/ris?id=10.1371%2Fjournal.pone.0115253"
         type="text/plain"
         profile="https://en.wikipedia.org/wiki/RIS_(file_format)"/>
</url>
\end{lstlisting}

\subsection{Signposting}
\label{Sec:Summary:Signposting}

Signposting is about communicating information in the Link header of a web resource associated with a scholarly digital object 
when HTTP-interacting with it. Listing \ref{lst:CrossRefmetalink} shows that CrossRef can use Signposting to support 
discovery of its metadata for a DOI-identified object.

\begin{lstlisting}[frame=single, caption={HTTP HEAD on a DOI provides a link to metadata describing the 
DOI-identified object}, label={lst:CrossRefmetalink}, float=*,
basicstyle=\ttfamily\scriptsize]
~\textbf{Request}~

curl -I http://dx.doi.org/10.1371/journal.pone.0115253

~\textbf{Response}~

  HTTP/1.1 303 See Other
  Vary: Accept
  Location: http://dx.plos.org/10.1371/journal.pone.0115253
  Link: <http://api.crossref.org/works/10.1371/journal.pone.0115253> 
   ; rel="describedby"
   ; type="application/json"
   ; profile="https://github.com/CrossRef/rest-api-doc"
  Content-Type: text/html;charset=utf-8
  Content-Length: 179
  Date: Thu, 05 May 2016 21:59:52 GMT
\end{lstlisting}

Publishers can also use Signposting to make metadata about their articles discoverable. 
Listing \ref{lst:PLOSmetalink} shows links from the \msg{Entry Page} to 
metadata available from the CrossRef API as well as two types of publisher metadata. Most importantly, it also shows links to 
all \msg{Publication Resources} as well as links that reveal the \msg{Identifying HTTP URI}. 
In general, publishers can use Signposting for each of their web resources associated with a scholarly object. 
As an example, Listing \ref{lst:PLOSpdf} shows an HTTP HEAD interaction with the PDF version of the PLOS article. 
The links provided in the HTTP response Link header characterize the resource as an article, 
convey its \msg{Identifying HTTP URI} (DOI at \msg{dx.doi.org}), 
and link back to the collection resource (the \msg{Entry Page}) from which 
other components of the scholarly object are linked. 

\begin{lstlisting}[frame=single, caption={HTTP HEAD on the Entry Page provides links to 
CrossRef and publisher metadata describing the DOI-identified object}, label={lst:PLOSmetalink}, float=*,
basicstyle=\ttfamily\scriptsize]
~\textbf{Request}~

curl -I http://journals.plos.org/plosone/article?id=10.1371/journal.pone.0115253

~\textbf{Response}~

  HTTP/1.1 200 OK
  Date: Mon, 09 May 2016 19:18:04 GMT
  Link: <info:eu-repo/semantics/article> 
   ; rel="type" , 
   <http://dx.doi.org/10.1371/journal.pone.0115253> 
   ; rel="persistent-id" ,
   <http://journals.plos.org/plosone/article/asset?id=10.1371%2Fjournal.pone.0115253.PDF> 
   ; rel="item"
   ; type="application/pdf" 
   ; sem-type="info:eu-repo/semantics/article"/> , 
   <http://journals.plos.org/plosone/article/asset?id=10.1371%2Fjournal.pone.0115253.XML> 
   ; rel="item"
   ; type="application/xml" 
   ; sem-type="info:eu-repo/semantics/article"/> , 
   <http://journals.plos.org/plosone/article/asset?unique&id=info:doi/10.1371/journal.pone.0115253.s012> 
   ; rel="item"
   ; type="text/html" 
   ; sem-type="info:eu-repo/semantics/objectFile"/> ,
   <http://api.crossref.org/works/10.1371/journal.pone.0115253> 
   ; rel="describedby"
   ; type="application/json"
   ; profile="https://github.com/CrossRef/rest-api-doc", 
   <http://journals.plos.org/plosone/article/citation/bibtex?id=10.1371%2Fjournal.pone.0115253>
   ; rel="describedby"     
   ; type="text/plain"
   ; profile="http://bibtex.org", 
   <http://journals.plos.org/plosone/article/citation/ris?id=10.1371%2Fjournal.pone.0115253>
   ; rel="describedby"     
   ; type="text/plain"
   ; profile="https://en.wikipedia.org/wiki/RIS_(file_format)", 
  Content-Type: text/html;charset=utf-8
  Content-Language: en-US
  Content-Length: 300137
\end{lstlisting}

\begin{lstlisting}[frame=single, label={lst:PLOSpdf}, caption={HTTP HEAD on the PDF article provides links to the DOI, 
the canonical URI, the collection resource, and characterizes the PDF as an article}, basicstyle=\ttfamily\scriptsize, float=*]
~\textbf{Request}~

curl -I http://journals.plos.org/plosone/article/asset?id=10.1371%2Fjournal.pone.0115253.PDF

~\textbf{Response}~

  HTTP/1.1 200 OK
  Date: Thu, 05 May 2016 21:59:52 GMT
  Last-Modified: Mon, 02 Mar 2015 22:52:02 GMT
  Content-Type: application/pdf
  Content-Length: 1794628
  Link: <info:eu-repo/semantics/article> 
   ; rel="type" , 
   <http://dx.doi.org/10.1371/journal.pone.0115253> 
   ; rel="persistent-id" ,
   <http://journals.plos.org/plosone/article?id=10.1371/journal.pone.0115253> 
   ; rel="collection"
   ; type="text/html"
   ; sem-type="info:eu-repo/semantics/article"
\end{lstlisting}

Note that Link headers could eventually become big if additional relations that serve other use cases would be included. 
While web servers and clients can deal with extensive HTTP headers, it may eventually become worthwhile to consider defining a relation type, for example 
\msg{links} dedicated to pointing at a document that contains links as an alternative for providing all links in the HTTP header.

\subsection{Recommendations}
\label{Sec:Summary:Recommendations}

Based on the above discussions, this section provides recommendations for 
CrossRef, academic publishers, and, more broadly, all operators of portals involved in scholarly communication. 

\subsubsection{Recommendations - CrossRef}
\label{Sec:Summary:Recommendations:CrossRef}

In order to support the Manifest and Bibliography requirements, CrossRef is invited to consider these recommendations:

\begin{itemize}
\item Recommendation 1 - Regarding Manifest: Implement ResourceSync Change Lists and/or Change Notifications to 
provide timely information regarding changed DOI registrations, i.e., new, updated, deleted. See Listing \ref{lst:CrossRefPLOS}. 
\item Recommendation 2 - Regarding Bibliography - Express ResourceSync Change Lists and/or Change Notifications in terms of 
the DOI at \msg{dx.doi.org} and provide a \msg{describedby} link pointing at the 
metadata that describes the DOI-identified object. See Listing \ref{lst:CrossRefPLOS}. 
\item Recommendation 3 - Regarding Bibliography: Implement Signposting as a means to support discovery of 
bibliographic metadata describing a DOI-identified object 
without the need for out-of-bound knowledge, by providing a \msg{describedby} link and associated attributes (as per Table \ref{tab:reltypes}) 
pointing from the DOI at \msg{dx.doi.org} to the metadata available via the CrossRef API. See Listing \ref{lst:CrossRefmetalink}.
\end{itemize} 

Note that both the CrossRef API and ResourceSync can support the Manifest requirement. However, regarding Manifest, 
the CrossRef API has drawbacks when compared to ResourceSync in that it involves recurrent searches against the 
entire database and returns both newly registered and changed DOIs without discrimination. In addition, the use of 
the same technology - ResourceSync - by CrossRef, publishers, institutional repositories, etc. 
would yield a considerable increase in interoperability for web-based scholarly communication. 

\subsubsection{Recommendations - Publishers \& Portals}
\label{Sec:Summary:Recommendations:Publishers}

In order to support the Manifest, Completeness, and Bibliography requirements, publishers are invited to consider these recommendations:

\begin{itemize}
\item Recommendation 4 - Regarding Manifest: Implement ResourceSync Change Lists and/or Change Notifications to provide timely information regarding new objects. 
See Listing \ref{lst:PLOSevent}.
\item Recommendation 5 - Regarding Manifest: Express ResourceSync Change Lists and/or Change Notifications for new objects in terms of 
the URI of the \msg{Entry Page} (e.g., the landing page, the HTML article), which is modeled as a collection. 
See Listing \ref{lst:PLOSevent}.
\item Recommendation 6 - Regarding Completeness: In ResourceSync Change Lists and/or Change Notifications, 
provide \msg{item} links and associated attributes (as per Table \ref{tab:reltypes}) that point from the \msg{Entry Page} (e.g., the landing page, the HTML article) 
to each 
of the \msg{Publication Resources}. See Listing \ref{lst:PLOSevent}.
\item Recommendation 7 - Regarding Completeness: For the \msg{Entry Page} (e.g., the landing page, the HTML article), 
implement Signposting by providing \msg{item} links and associated attributes (as per Table \ref{tab:reltypes}) 
that point at each of the \msg{Publication Resources}. See Listing \ref{lst:PLOSmetalink}
\item Recommendation 8 - Regarding Completeness: For each \msg{Publication Resource}, implement Signposting 
by providing a \msg{collection} link and associated attributes (as per Table \ref{tab:reltypes}) pointing at the \msg{Entry Page}. 
See Listings \ref{lst:PLOSpdf}. 
\item Recommendation 9 - Regarding Bibliography: In ResourceSync Change Lists and/or Change Notifications, 
provide \msg{describedby} links and associated attributes (as per Table \ref{tab:reltypes}) that point at bibliographic 
metadata describing the object, including the metadata available from the CrossRef API. 
See Listing \ref{lst:PLOSevent}.
\item Recommendation 10 - Regarding Bibliography: In ResourceSync Change Lists and/or Change Notifications, 
include the \msg{persistent-id} link ponting to the \msg{Identifying HTTP URI} (e.g., DOI at \msg{dx.doi.org}), 
if one exists. See Listing \ref{lst:PLOSevent}.
\item Recommendation 11 - Regarding Bibliography: For the \msg{Entry Page} (e.g., the landing page, the HTML article), implement 
Signposting by providing \msg{describedby} links and associated attributes (as per Table \ref{tab:reltypes}) that point at bibliographic metadata 
describing the object, including the metadata available from the CrossRef API. See Listing \ref{lst:PLOSmetalink}. From publisher metadata resources, 
link back to the \msg{Entry Page} using the \msg{describes} relation type.
\item Recommendation 12 - Regarding Bibliography: For the \msg{Entry Page} and all \msg{Publication Resources}, 
implement Signposting by providing a \msg{persistent-id} link ponting to the \msg{Identifying HTTP URI} (e.g., DOI at \msg{dx.doi.org}), 
if one exists. See Listings \ref{lst:PLOSmetalink} and \ref{lst:PLOSpdf}.
\end{itemize}

\section{Conclusion}
\label{Sec:Conclusion}

Ingest is a major contributor to the cost of preserving e-journals.
Cost is the major reason that archives contain less than half the content
they should~\cite{HalfEmptyArchive}. This paper has shown that the use of existing infrastructure (CrossRef API)  
combined with the introduction of new infrastructure (ResourceSync Change Lists and/or Change Notifications, Signposting) 
promises significant cost savings, and the potential for improvement in the quality of both the completeness
of archived content, and of the metadata (including identity) describing it. The implementation of new infrastructure certainly comes at a cost. 
But, 
software that implements ResourceSync Change Notifications is available\footnote{\url{https://github.com/resync/resourcesync_push}} and 
individual notifications can easily be grouped into Change Lists. Also, while the Link headers used for Signposting may come across as 
intimidating, in many cases, they can be automatically generated using web server rewrite rules because publisher 
platforms typically have a consistent approach regarding URI patterns for web resources that are part of an object. In other cases, typed links 
can be implemented using internal API calls. 

The combination of ResourceSync implemented by CrossRef and Signposting implemented by publishers 
would go a long way to address crucial problems faced by current archiving efforts, 
when DOI-identified objects are concerned. For publishers that assign DOIs, implementing 
ResourceSync would also be beneficial because it would help address the asynchronous registration of a DOI and 
the publication of the DOI-identified object, and would allow monitoring whether newly published articles are also 
made available for archiving by file transfer. For publishers that don't assign DOIs, and hence can't rely on 
Change feeds from CrossRef to address the Manifest requirement, implementation of ResourceSync is essential 
to allow (archiving) applications to remain up to date with evolving content. Generally, implementation of ResourceSync by publishers, 
repositories, etc. would significantly augment interoperability for web-based scholarly communication.

Because the ResourceSync and Signposting guidelines apply to the URIs at dx.doi.org, 
and to the targets they redirect to, CrossRef could also build a tool that audits publishers' compliance.
The carrot could be, say, lower dues for conformance, and the stick could be,
say, naming-and-shaming non-compliant publishers.
Most e-journals are now published by one of a limited number of publishing
platforms, which would have to make the necessary changes. 
They would be driven by demand from their customer publishers.

The proposed infrastructure would also benefit other applications. 
For example, although DOIs have been widely adopted,
research has shown that their full potential benefits are not being achieved.
The fundamental benefit DOIs should provide is as a persistent identifier,
but:
\begin{quote}
we found a significant number of references to papers by their locating URI
instead of their identifying URI.
For these links, the persistence intended by the DOI persistent identifier infrastructure
was not achieved.~\cite{vdSompel-WWW2016}
\end{quote}
Locating URIs are used for many reasons,
but a key one is that they, and not the DOI URI, are available to be cut-and-pasted 
in the usual way from the browser's address bar. The \msg{persistent-id} link provided 
by Signposting and ResourceSync would supply 
the information destroyed by the use of locating URIs rather than DOI URIs, 
if publishers could be persuaded to include it in the HTTP response header of web resources 
that are part of a DOI-identified object. Tools like citation managers, browsers, 
search engines could evolve to use this link to record the \msg{Identifying HTTP URI} instead 
of the location URI of the web resource. Adding this link would also allow connecting open 
annotations\footnote{\url{https://hypothes.is/annotating-all-knowledge/}} made to an HTML or PDF 
version of a paper back to the DOI. More generally, it would uniformly provide the HTTP version of the DOI, 
in contrast to current practice in citations, landing pages, and bibliographic records that still frequently 
use a string rather than URI version of the DOI, and, many times even use a publisher HTTP URI instead of the 
DOI HTTP URI. Hence, we believe it would be in CrossRef's 
interest to encourage their members to adopt Signposting and ResourceSync, as described above.

The low-barrier Signposting approach as such has enormous potential for web-based scholarly communication. 
For example, the \msg{describedby} links potentially allow the creation of a citation graph by crawling, 
assuming article reference lists are openly accessible with each reference linking to 
an HTTP URIs that, via Signposting, leads to structured bibliographic metadata about the referenced article. If, additionally, 
\msg{author} links would be provided, ideally pointing to the ORCID at \msg{orcid.org}, co-author graphs could be created by crawling. 
As we explore Signposting beyond the pattern described in this paper, information will be made available at \url{http://signposting.org}. 
What are you waiting for? 

\subsection*{Acknowledgments}

Grateful thanks are due to
Fernando García-Loygorri,
Daniel Vargas,
Craig Van Dyck,
Victoria Reich, 
Simeon Warner,
Geoff Bilder, 
Harihar Shankar, 
and Shawn Jones.

\bibliographystyle{plain}
\bibliography{DoiBasedIngest.bib}

\begin{thebibliography}{10}

\bibitem{Brook-11-2014}
Michelle Brook, Peter Murray-Rust, and Charles Oppenheim.
\newblock {The Social, Political and Legal Aspects of Text and Data Mining
  (TDM)}.
\newblock {\em {D-Lib Magazine}}, 20(11/12), November 2014.
\newblock \url{http://dx.doi.org/10.1045/november14-brook}.

\bibitem{CLOCKSS}
{CLOCKSS}.
\newblock {CLOCKSS: A Trusted Community-Governed Archive}.
\newblock \url{http://www.clockss.org/}.

\bibitem{CrossRefApi}
{CrossRef}.
\newblock {CrossRef REST API}.
\newblock
  \url{https://github.com/CrossRef/rest-api-doc/blob/master/rest_api.md}.

\bibitem{CrossRef}
{CrossRef}.
\newblock {crossref.org}.
\newblock \url{http://www.crossref.org/}.

\bibitem{sitemap}
{Google Inc., Yahoo Inc., and Microsoft Corporation}.
\newblock {Sitemap protocol}.
\newblock \url{http://sitemaps.org}.

\bibitem{IanaLinkRelationRegistry}
{IANA}.
\newblock {Link Relations}.
\newblock
  \url{https://www.iana.org/assignments/link-relations/link-relations.xml}.

\bibitem{EvanescentWeb}
Martin Klein, Herbert Van~de Sompel, Robert Sanderson, Harihar Shankar,
  Lyudmila Balakireva, Ke~Zhou, and Richard Tobin.
\newblock {Scholarly Context Not Found: One in Five Articles Suffers from
  Reference Rot}.
\newblock {\em {PLoS One}}, December 2014.
\newblock \url{http://dx.doi.org/10.1371/journal.pone.0115253}.

\bibitem{OAI-PMH}
Carl Lagoze, Herbert {Van de Sompel}, Michael Nelson, and Simeon Warner.
\newblock {The Open Archives Initiative Protocol for Metadata Harvesting}.
\newblock \url{http://www.openarchives.org/OAI/openarchivesprotocol.html},
  2002.

\bibitem{LOCKSS-MetadataDatabase}
LOCKSS.
\newblock {{LOCKSS: Metadata Database}}.
\newblock
  \url{http://documents.clockss.org/index.php/LOCKSS:_Metadata_Database}, July
  2014.

\bibitem{LOCKSS-MetadataExtractor}
LOCKSS.
\newblock {LOCKSS: Metadata Extractor}.
\newblock
  \url{http://documents.clockss.org/index.php/LOCKSS:_Extracting_Bibliographic_Metadata},
  July 2014.

\bibitem{ResourceSync-spec}
{NISO and The Open Archives Initiative}.
\newblock {Z39.99-2014, ResourceSync Framework Specification}.
\newblock \url{http://www.openarchives.org/rs/toc}, 2014.

\bibitem{RFC5988}
Mark Nottingham.
\newblock {RFC 5988}: {Web Linking}.
\newblock \url{http://tools.ietf.org/html/rfc5988}, October 2010.

\bibitem{link-hints}
Mark Nottingham.
\newblock {Link Hints Internet Draft}.
\newblock \url{https://tools.ietf.org/html/draft-nottingham-link-hint-00}, June
  2013.

\bibitem{UCPortico}
{Preservation CKG}.
\newblock {Reliability of Portico for preservation of UC’s e-journals}.
\newblock
  \url{https://wiki.library.ucsf.edu/download/attachments/328601907/portico_report_2014_11_03.docx},
  November 2014.

\bibitem{LOCKSS}
LOCKSS Program.
\newblock {Lots Of Copies Keep Stuff Safe}.
\newblock \url{http://lockss.org/}.

\bibitem{RosenthalMetadataIIPC2013}
David S.~H. Rosenthal.
\newblock {Talk on LOCKSS Metadata Extraction at IIPC 2013}.
\newblock
  \url{http://blog.dshr.org/2013/04/talk-on-lockss-metadata-extraction-at.html},
  April 2013.

\bibitem{HalfEmptyArchive}
David S.~H. Rosenthal.
\newblock {The Half-Empty Archive}.
\newblock \url{http://blog.dshr.org/2014/03/the-half-empty-archive.html}, March
  2014.

\bibitem{CLOCKSS-Audit}
David S.~H. Rosenthal.
\newblock {TRAC Certification of the CLOCKSS Archive}.
\newblock
  \url{http://blog.dshr.org/2014/07/trac-certification-of-clockss-archive.html},
  July 2014.

\bibitem{PorticoWebsite}
ITHAKA S+R.
\newblock {PORTICO: Your content. Preserved here.}
\newblock \url{http://www.portico.org/digital-preservation/}.

\bibitem{rsync}
Andrew Tridgell and Paul Mackerras.
\newblock The rsync algorithm.
\newblock Technical Report TR-CS-96-05, Australian National University, 1996.

\bibitem{vdSompel-WWW2016}
Herbert Van~de Sompel, Martin Klein, and Shawn~M Jones.
\newblock {Persistent URIs Must Be Used To Be Persistent}.
\newblock In {\em {25th international world wide web conference}}, April 2016.
\newblock \url{http://arxiv.org/abs/1602.09102}.

\bibitem{vdSompel2015}
Herbert {Van de Sompel} and Michael Nelson.
\newblock {Reminiscing About 15 Years of Interoperability Efforts}.
\newblock {\em {D-Lib Magazine}}, 21(11/12), November 2015.
\newblock \url{http://dx.doi.org/10.1045/november2015-vandesompel}.

\bibitem{StmReport2015}
Mark Ware and Michael Mabe.
\newblock {The STM Report: An overview of scientific and scholarly journal
  publishing}.
\newblock Technical Report 4th Edition, International Association of
  Scientific, Technical and Medical Publishers, March 2015.
\newblock \url{http://www.stm-assoc.org/2015_02_20_STM_Report_2015.pdf}.

\end{thebibliography}

\end{document}